# In-situ neutron diffraction during reversible deuterium loading in Ti-rich and Mn-substituted $Ti(Fe,Mn)_{0.90}$ alloys

Erika Michela Dematteis,[a,b)*] Jussara Barale,[a),§§] Giovanni Capurso,[c),§] Stefano Deledda,[d)] Magnus H. Sørby,[d)] Fermin Cuevas,[b)] Michel Latroche,[b),†] and Marcello Baricco[a)]

[a)] Department of Chemistry, Inter-departmental Center NIS and INSTM, University of Turin, Via Pietro Giuria 7, 10125 Torino, Italy

[b)] Univ Paris Est Creteil, CNRS, ICMPE, UMR 7182, 2 rue Henri Dunant, 94320 Thiais, France

[c)] Institute of Hydrogen Technology, Helmholtz-Zentrum Hereon, Max-Planck-Straße 1, 21502 Geesthacht, Germany

[d)] Department for Hydrogen Technology, Institute for Energy Technology, P.O. Box 40, Kjeller, NO-2027, Norway

*Corresponding author: Erika Michela Dematteis

E-mail address: erikamichela.dematteis@unito.it

[†]Deceased author

[§]Present address: Polytechnic Department of Engineering and Architecture, University of Udine, via del Cotonificio 108, 33100 Udine, Italy

[§§]Present address: Tecnodelta srl, Via Francesco Parigi, 5H, 10034, Chivasso, Italy

**Graphical Abstract**

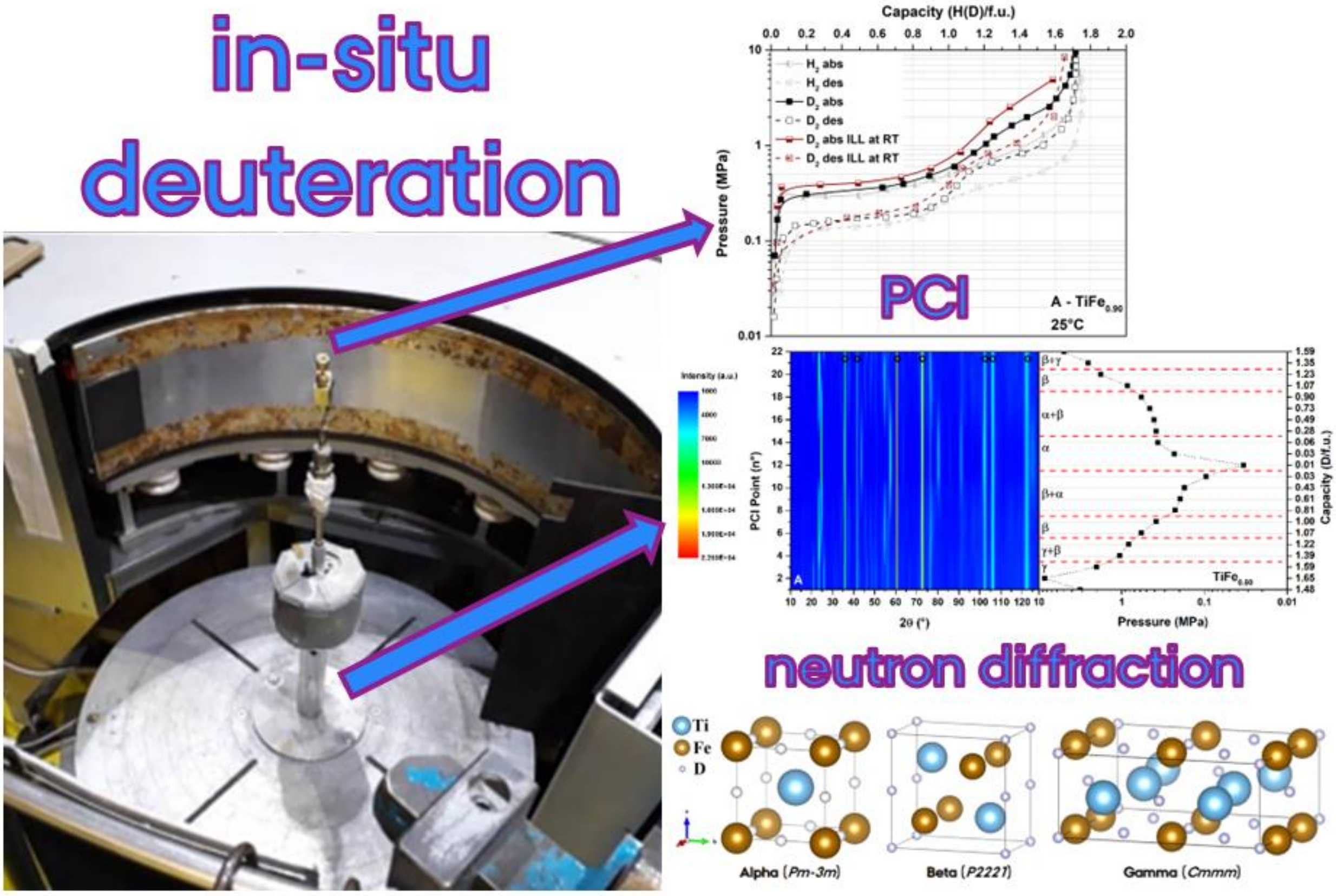

**Abstract**

Hydrogen is an efficient energy carrier that can be produced from renewable sources, enabling the transition towards $CO_2$-free energy. Hydrogen can be stored for a long period in the solid-state, with suitable alloys. Ti-rich $TiFe_{0.90}$ compound exhibits a mild activation process for the first hydrogenation, and $Ti(Fe,Mn)_{0.90}$ substituted alloys can lead to the fine tuning of equilibrium pressure as a function of the final application.

In this study, the crystal structure of $TiFe_{(0.90-x)}Mn_x$ alloys ($x$ = 0, 0.05 and 0.10) and their deuterides has been determined by *in-situ* neutron diffraction, while recording Pressure-Composition Isotherms at room temperature. The investigation aims at analysing the influence of Mn for Fe substitution in Ti-rich $Ti(Fe,Mn)_{0.90}$ alloys on structural properties during reversible deuterium loading, which is still unsolved and seldom explored. After activation, samples have been transferred into custom-made stainless-steel and aluminium alloy cells used for *in-situ* neutron diffraction experiments during deuterium loading at ILL and ISIS neutron facilities, respectively. The study enables remarkable understanding on hydrogen storage, basic structural knowledge, and support to the industrial application of TiFe-type alloys for integrated hydrogen tank in energy storage systems by determining the volume expansion during deuteration. Furthermore, the study demonstrates that different contents of Mn do not significantly change the volumetric expansion during phase transitions, affecting only the deuterium content for the γ phase and the cell evolution for the β phase. The study confirms that the deuterated structures of the γ phase upon absorption, β and α phase upon desorption, correspond to S.G. *Cmmm*, $P222_1$ and *Pm*-3*m*, respectively.

## 1. Introduction

The increase of pollution and global temperatures reveals how urgent and essential is to look for alternatives to fossil fuels. The efficient storage of renewable energy would enable the transition towards $CO_2$-free energy. Hydrogen can be produced from water electrolysis using renewable or nuclear energy sources and, as an efficient energy carrier, it can be stored for a long period. Metal hydrides are promising and safe materials for solid-state hydrogen storage, allowing storing hydrogen in mild conditions (*i.e.* ambient pressure and close to room temperature – RT) with high volumetric densities.[1–4]

TiFe intermetallic is a low-cost and promising compound for $H_2$ storage.[5] TiFe exhibits, however, drawbacks and particular features as hydrogen storage material. Firstly, it is difficult to activate towards hydrogen absorption, *i.e.*, to promote hydrogen uptake for the first time. A thermal cycling between RT and 400 °C is necessary for activation and the full capacity is only attained after tens of hydrogen sorption cycles.[6] Secondly, differing from the paradigmatic $LaNi_5$ alloy, Pressure-Composition-Isotherms (PCI) are characterized by two plateau pressures, instead of showing a single one.[6] Thirdly, TiFe was reported to have significant isotope effect.[7,8] Near RT, plateau pressures of deuterides have been reported significantly higher than those of the hydrides, which makes it difficult to achieve a full deuteration of TiFe alloy for pressures below 10 MPa.[7] All these properties are at the origin of controversial results concerning the structure of TiFe hydrides/deuterides.

TiFe alloy shows a cubic CsCl-type (B2, space group S.G. *Pm*-3*m*; 1:1 atomic composition, $a$ = 2.975 Å)[9,10] crystal structure, as called α phase (**Table 1**).[10–19] During hydrogenation, the α phase can accommodate up to 0.10 hydrogen atoms per formula unit (H/f.u.) as solid solution.[6] The lattice expansion in the solid solution region was described by Reilly *et al.* back in 1980, defining a α/β phase boundary at $TiFeH_{0.04}$.[14]

The dissolution of hydrogen in the α phase is observed as an increasing branch in PCI curves at low pressures, followed by a low pressure plateau that characterises the formation of the β phase, a monohydride with formula TiFeH.[12]

Through the low-pressure plateau, the orthorhombic β phase forms (*S.G. P*$222_1$),[10–12,16,20,21] even if discrepancies concerning the TiFe hydride crystal structures are still present in the literature (**Table 1**).[13,18,19] Firstly, in 1974, Reilly *et al.* reported a tetragonal structure ($TiFeH_{1.06}$, *a* = 3.18 Å, *c* = 8.73 Å) determined by XRD.[6] Furthermore, in the literature, two different β phases have been reported, namely $\beta_1$ and $\beta_2$, related to the TiFeH and $TiFeH_{1.4}$ compositions, respectively.[21–23] With further proceeding of the hydrogenation reaction, at higher hydrogen pressure, a second plateau is observed in PCI curves, related to the full hydrogenation of the material up to a $TiFeH_2$ composition (γ phase). As it can be observed from **Table 1**, structural studies of the γ phase suggested different possible structures, *i.e.* cubic ($TiFeH_{1.93}$, *a* = 6.61 Å),[6] orthorhombic (*S.G. Cmmm*)[11,15,16,18,24,25] and monoclinic structure (*S.G. P2/m*).[13,15,19,21,24,26,27] In the monoclinic structure, deuterium is localised in four special positions with octahedral coordination.[27]

The structural evolution of TiFe upon deuterium loading was described by Schafer *et al.* back in 1980.[27] From the disordered α phase the D-ordering on octahedral sites is taking place with an orthorhombic distortion to form the ordered β phase, then the distortion further proceeds to form the γ phase, with two additional deuterium positions close to Fe atoms towards the monoclinic structure.[27] Indeed, distortions of peak shape have been evidenced in neutron diffraction studies, correlated to the formation of possible defects like dislocations, vacancies, disorder, strain and distortions.[3,12,28]

Substitutional effects on hydrogenation properties in TiFe have been recently reviewed by Dematteis *et al.*[29], TiFe can accommodate by substitution different transition metals and *p*-block elements on either Ti or Fe sites.[3,28]

Mn-substituted TiFe materials are good candidate for $H_2$ storage with the di-hydride having a gravimetric capacity of 1.9 wt.% $H_2$, mild activation conditions and enhanced hydrogenation properties.[1,2] Multi-substituted alloys can display enhanced activation properties and the possibility of tuning plateau pressure as a function of substituent concentration in the formulated alloy, as in the case of $TiFe_{0.90-x-y}Mn_xCu_y$-substituted materials, that enable hydrogen storage in milder conditions of pressure and temperature.[30] When Mn is introduced in the alloy, it substitutes Fe enlarging the cell parameter, as reported by Fruchart *et al.* using neutron diffraction for the $TiFe_{0.90}Mn_{0.10}$ composition ($a$ = 2.985 Å).[19] Upon hydrogenation, Mn-substituted alloy is reported to crystallize in an orthorhombic structure, that was assigned to *S.G. P2cm* or *Pmc2₁* for $TiFe_{0.90}Mn_{0.10}H_{1.25}$.[19] for the γ phase, a monoclinic structure has been reported in Mn-substituted materials[19,23] and it has been recently studied by *in-situ* synchrotron radiation studies on $TiFe_{0.80}Mn_{0.20}$.[31]

In real applications, the selection of a proper alloy with adequate thermodynamics in the operative temperature and pressure ranges is a milestone. Costs, use of non-critical raw materials and high hydrogen content should be also considered. Furthermore, the alloy should have an easy activation procedure and fast hydrogen sorption kinetics to be integrated in the plant.[32] Mn-substituted TiFe alloys fulfil these criteria. As an example of real applications, the EU-funded HyCARE project aims at realizing an innovative large-scale hydrogen storage tank based on TiFe-based solid-state carrier.[33,34] One of the goals is to investigate and select the composition and structure of an optimum alloy for large-scale hydrogen storage at the solid-state.[35–38] The selected material will be scaled up to approx. 5 tons and used in an efficient integrated energy storage system, where about 40 kg of hydrogen will be safely stored. The study of the industrially produced alloy has been recently published discussing its hydrogenation properties and activation procedure compared with the lab-scaled one.[39]

The present investigation, carried out in the framework of the HyCARE project, is an *in-situ* neutron diffraction study of the structural evolution of three different Mn-substituted TiFe alloys to better understand the evolution of their crystal structures upon hydrogenation.

Using deuterium instead of hydrogen will guarantee the avoidance of relatively high H incoherent scattering, while the combination of neutron diffraction as well as in-lab x-ray diffraction will allow to determining crystal structures and composition, as well as Ti and Mn substitutions since Ti/Fe contrast in ND-experiments is high (bound coherent scattering length, $b_c$ = -3.44 and 9.45 fm for Ti and Fe, respectively. However, care should be taken to analyse simultaneous occupation of Ti and Mn atoms on same site as they have similar scattering lengths ($b_c$ (Mn) = -3.73 fm).

The determination of structural volumetric expansion during the process supports the scale-up and modelling of the hydrogen storage tank from an engineering point of view.

Finally, the aim of the paper is to analyse the influence of stoichiometry and Mn for Fe substitution in TiFe-type alloys on structural properties during reversible deuterium loading, which is a topic still unsolved and seldom explored. The $TiFe_{(0.90-x)}Mn_x$ alloys (x = 0, 0.05 and 0.10) have been selected within a promising composition range for an integrated, renewable, large scale stationary hydrogen storage tank [40].

## 2. Experimental

### 2.1. Sample preparation

In the frame of the HyCARE project, TiFe-type alloys have been synthesized in an induction furnace, annealed at 1000 °C for 1 week and subsequently activated and deuterated up to the γ phase, nominally $TiFeD_2$.[40] Three of these alloys (Ti-rich samples s2, s6, s9 in ref.[40], now referred to as A: $TiFe_{0.90}$, B: $TiFe_{0.85}Mn_{0.05}$, and C: $TiFe_{0.80}Mn_{0.10}$, respectively) were selected, owing to improved hydrogenation properties, to be structurally characterised through *in-situ* neutron diffraction (ND), and to determine their phase composition, structural evolution during sorption of deuterium at RT and the influence of Mn content on structural and deuteration properties. For each $TiFe_{(0.90-x)}Mn_x$ sample ($x$ = 0, 0.05, 0.10) the formula unit (f.u.) is expressed as $Ti_{1.05}Fe_{(0.95-x)}Mn_x$ that, considering the occupancy of crystallographic sites, corresponds to [1*a*][1*b*]:[$Ti_{0.05}Fe_{(0.95-x)}Mn_x$][Ti]. Argon-filled glove boxes, with a circulation purifier, were used to prepare, store, and manipulate samples. $O_2$ and $H_2O$ levels were lower than 1 ppm to minimize oxidation and degradation of the materials.

### 2.2. Characterizations

Previous characterizations evidenced that samples at the Ti-rich side have good storage reversible capacities and, after synthesis and annealing, they present a microstructure with a TiFe-substituted matrix and secondary phases precipitates at grain boundaries (assigned to Ti-rich β-Ti and $Ti_4Fe_2O$-type phases).[40] The compositional balance determined by EPMA (electron probe microanalysis) analysis suggests that both Ti and Mn partially substitute Fe. The formation of a limited amount of secondary phases, reactive to hydrogen, allows first hydrogenation in mild conditions of temperature and pressure. Finally, Mn-substitution has been evidenced to enlarge TiFe cell volume, reduce secondary phases amount, reduce equilibrium pressure, reduce hysteresis and increase reversible capacity in mild conditions.[40]

### *2.2.1. Deuteration properties*

Activation, kinetic studies and PCI curves were monitored with a home-made Sieverts-type apparatus under hydrogen, as reported in ref.[40].

After the activation, the entire synthetized sample (approximately 8-10 g) was loaded into a custom stainless steel sample cell. PCI curves were first measured in lab under deuterium gas, both in absorption and desorption at 25 °C by the manometric Sieverts' method, using calibrated and thermalized volumes. A water thermostatic bath was used for maintaining a constant temperature during PCI recording.

### *2.2.2. In-situ neutron diffraction*

Powder ND-patterns were obtained at RT at Institut Laue-Langevin (ILL) on beamline D1B (λ=1.286 Å, Cycle 20201, Proposal CRG 5-22-771 [41]) in a 2θ-range of 10–128°, step size 0.1° and equipped with $3He/CF_4$ position sensitive detector composed of a system of multi electrodes with 1280 wires. Samples were deuterated at *ca.* 8-9 MPa and transferred to the neutron facility for *in-situ* neutron diffraction studies already inside the stainless-steel cell, which were placed in the beam and connected to a manual Sieverts-type volumetric rig. ND-patterns were recorded in the as-mounted state as a function of deuterium pressure. At the beginning of experiments, a high deuterium pressure was applied (ca. up to 9 MPa) to guarantee a complete deuteration to the γ phase. The deuterium uptake was determined using a pressure gauge. ND patterns were recorded every 10 minutes during the deuteration process. Several patterns were then collected and summed up, once thermodynamic equilibrium was attained.[41] By stepwise lowering the deuterium pressure, ND patterns were collected during acquisition of PCI desorption isotherm at RT, followed by the same analysis on absorption increasing the deuterium pressure. In the present study, fully deuterium-free samples were not analysed, since the lower pressure used was 0.02 MPa and only evacuation at RT was performed; thus, the α phase always contains a small amount of deuterium as solid solution. D1B experiment allowed to characterize various phase transitions. To fully solve the structure of the γ-dihydride, a single scan at high resolution beamline D2B has been performed (λ=1.594 Å, Proposal 83988 EASY-593), in a 2θ-range of 10–160° with step size 0.05°.[42]

All neutron diffraction patterns obtained at ILL were refined by the Rietveld method,[43] using the FullProf package.[44] The pseudo-Voigt profile function was used to model the shape of diffraction peaks and the instrumental function was obtained by analysing a Si standard.

In addition, ND patterns were also obtained at RT at the Science and Technology Facilities Council (STFC) ISIS Neutron and Muon Source on GEM beamline (Experiment number RB192055).[45]

ISIS samples were desorbed under dynamic vacuum at 200 °C for 30 minutes and transferred to the neutron facility for *in-situ* neutron diffraction studies. The un-deuterated sample was loaded at ISIS using an Ar-filled glovebox into an Al cylindrical cell that can reach 300 bar of pressure and 100 °C (A2 SI-4105-409).

The samples were placed in the beam and connected to a manual Sieverts-type volumetric rig, where volumes were calibrated using a reference volume and helium. ND-patterns of the samples were recorded in the as-mounted state and the beam delimiter was optimised to 30x10 mm (HxW). A long acquisition on as-mounted sample was performed before absorption.

Then, the buffer volume was filled with increasing deuterium pressure, short acquisitions were started and the valve to the cell was opened during the first acquisition. Deuterium loading was monitored until the equilibrium conditions were reached. When changes were no longer observed, a long acquisition was recorded.

ND patterns were recorded every 10 minutes during the deuteration process, while 60 minutes-long acquisition were recorded at the equilibrium, both in absorption and desorption. Pressure and PCI curves were continually monitored.

All data sets obtained at ISIS from long acquisitions were analysed with the Rietveld method[43], using the software TOPAS Academic version 6.[46] Data from all six detector banks were used simultaneously in each refinement. The instrumental function was given by macros developed for GEM based on a pseudo-Voigt function convoluted with the Ikeda-Carpenter function.[47]

## 3. Results and Discussion

### *3.1. Pressure Composition Isotherms and isotopic effects*

PCI curves under deuterium were obtained at RT at ILL, and then compared with those under hydrogen previously determined in lab at 25 °C.[40,48] As it can be observed from **Figure 1**, lab PCIs under deuterium show similar capacities and hysteresis, as well as similar slopes, compared to those monitored under hydrogen. While significant isotopic effects were previously reported in the literature for TiFe,[21] the present investigation for Ti-rich Ti(Fe,Mn) alloys evidences minor isotopic effects for the low-pressure plateau. This might tentatively be attributed to their different Ti-contents. However, the high-pressure plateau is shifted towards higher equilibrium pressures for deuterium, as compared to hydrogen for all three samples characterized. *In-situ* evaluations of PCI curves at RT during deuterium loading at ILL are reproducing well the deuterated PCI curves at 25 °C monitored in lab, considering differences in temperature and possible experimental faults due to different setups.

### *3.2. In-situ neutron diffraction experiments*

#### *3.2.1 Phases evolution and transition*

The *in-situ* diffraction analyses performed at ILL during deuterium absorption and desorption are outlined for samples A, B and C in **Figure 2**, together with the monitored PCI curves. Phase formation and their transition can be appreciated following the peak intensity lines in the two-dimensional contour plot displayed on left in **Figure 2**.

The analyses allowed the phase identification and evolution of their crystal structure (solid solution, two-phase reaction domains and single-phase deuterated phases) during reversible deuterium sorption.

As it can be observed **Figure 2-A**, the single γ phase upon absorption for the $TiFe_{0.90}$ sample has been obtained at high pressure (approx. 9 MPa $D_2$), with a final deuterium capacity of 1.65 D/f.u. according to PCI data. During desorption, the single-phase β is detected at 1.00-1.07 D/f.u., while it is detected between 1.07-1.23 D/f.u. during absorption. Single α phase was detected below 0.06 D/f.u. deuterium capacity. The recorded capacities reported in PCI are expected to be higher than deuterium occupancy obtained during the *in-situ* analysis by Rietveld refinement, as it can be observed in **Figure S1**, since deuteride-forming secondary phases such as β-Ti and $Ti_4Fe_2O$ are detected in the samples. Nevertheless, owing to the low amount of secondary phases, differences between the deuterium content determined by PCI or refinement are within experimental errors.

**Figure 2-B** displays the formation of phases and their transitions as a function of pressure for the $TiFe_{0.85}Mn_{0.05}$ sample. The formation of the single γ phase upon absorption occurs at high pressures (approx. 9 MPa $D_2$) with a 1.89 D/f.u. final deuterium capacity, according to PCI data. During desorption, a single-phase β is detected at 1.11 D/f.u., while the capacity related to single-phase β is equal to 1.28 D/f.u. during absorption. Single α phase was detected below 0.18 D/f.u. deuterium content.

Finally, **Figure 2-C** evidences the formation of the single γ phase for the $TiFe_{0.80}Mn_{0.10}$ sample upon absorption at high pressure (approx. 9 MPa $D_2$) with a final 1.99 D/f.u. deuterium content. The single-phase β was recorded at 1.12-1.27 D/f.u. during desorption. Single α phase was detected below 0.24 D/f.u. deuterium content.

**Figure 3** shows the phase evolution during an absorption and desorption cycle for Sample C measured at the GEM time-of-flight instrument at ISIS. The same phase transitions observed during the measurement at ILL (**Figure 2**) were observed. However, some notable differences can be pointed out. The sorption cycle started from the deuterium-free alloy (α phase) and proceeded with absorption up to 9.8 MPa. As expected, deuteride phases are formed upon increasing the pressure. First, the β phase was observed above 0.2 MPa, followed by the formation of the γ phase above 0.3 MPa, in agreement with PCI data shown in **Figure 1**. It is worth noting that a complete transformation from α to β and from β to γ was not observed and still 8.9 wt.% of α and 3.7 wt.% of β phases were still detected at 9.8 MPa (as quantitatively determined by Rietveld refinement). Accordingly, the deuterium capacity reported in **Figure 3** (D/f.u. = 1.7 at 9.8 MPa), which refers to the overall deuterium content of the material, is lower than that measured at ILL (D/f.u. = 1.99 at 8.37 MPa, **Figure 2**). This could be due to an incomplete activation of sample C prior to the measurement at ISIS, which limited deuterium absorption. As the pressure was decreased, desorption set in. The γ phase was no longer detected below 0.3 MPa and only the α phase with a residual D/f.u. = 0.03 was observed below 0.05 MPa.

### *3.2.2 Rietveld refinement*

The Rietveld refinement was performed on the experimental patterns to determine the crystal structures and phases composition (and related deuterium content in the crystal structure, **Table 2**). Ti/Fe contrast in ND-experiments, where the bound coherent scattering length is $b_c$ = -3.44 and 9.45 fm for Ti and Fe, respectively, is rather high. However, care should be taken to analyze the simultaneous occupancy of Ti and Mn atoms at the same site, as they have similar scattering lengths ($b_c$ (Mn) = -3.73 fm). For this reason, the site occupancy factors (SOF) of Ti, Fe and Mn have been fixed equal to that determined by EPMA analysis performed in ref.[40] and reported in **Table 3**, while deuterium occupancies at different sites were refined through the sequential *in-situ* analysis.[49] Since the compositions of analyzed samples are located on the Ti-rich side of the Ti(Fe,Mn) phase field, results confirm the Ti substitution on Fe site. Challet *et al.* already demonstrated that the combination of in-lab XRD, EPMA chemical analysis and density measurements can allow to establish location of possible vacancies, as well as Ti and Mn substitutions.[50] In a recent review on TiFe-substituted compounds,[29] the authors collected lattice parameters and Mn content for different alloys in the literature. Sometimes it is not possible to directly compare the lattice parameters with the real amount of Mn in the TiFe-phase, however, the data determined in this study are in line with literature determinations, supporting the declared concentration of Mn in the sample. Atomic positions and isotropic displacement factors were refined and assessed for single-phase Ti(Fe,Mn)-D patterns and then fixed during the sequential refinement, because of the limited resolution of D1B beamline. Isotropic displacement factors were observed to vary between 0.1-0.5 $Å^2$ for metals and 1.8-2.0 $Å^2$ for deuterium. The shape factor of the pseudo-Voigt function was refined through the sequential refinement, since it can change because of strain generated during absorption and desorption. The asymmetry factor at low angles was refined on single-phase α patterns and then maintained fixed for all refinements. It is worth noting that during the *in-situ* analysis, anisotropic peak broadening effects have been observed: in fact, in the β phase, asymmetric profiles are present on (011) and (100) peaks, while for the γ phase, asymmetry in (200) and (020) peaks in opposite angle directions was recorded, as it can be observed in published research datasets and refinements (see the Supplementary data section and ref.[49]), and, as an example, in the zoom on Rietveld refinement reported in **Figure S2**. This result reflects a high strain and the presence of micro-twins and dislocations in the above-mentioned samples during sorption, as already reported in previous literature studies.[12] These effects can be solved by implementing different strain models, but this was considered to be outside of the scope of this work. As a matter of fact, the non-implementation of these asymmetries results in higher Rietveld R-factor for the β and γ phases compared to the α phase, as it can be observed in **Table 2**.

All samples contain a small amount of fluorite-type $TiH_2$ phase, in which, according to the ternary Ti-Fe-Mn diagram,[51] Ti is partially substituted by Fe and Mn. The precipitation of this phase results from the fact that the Ti-content of the alloys (52.5 at.%) is at the limit of the homogeneity domain of the CsCl-type Ti(Fe,Mn) phase. In addition, only at low deuterium pressures, the presence of an oxide $Ti_4Fe_2O$-type phase could be detected and refined, while it could not be distinguished at high deuterium pressures, due to severe peak overlapping. Secondary phases have been reported to have a major role in improving activation at milder conditions of temperature and pressure with respect to the pure compound, since they can interact with hydrogen and behave as hydrogenation gates during the first hydrogenation of the alloy.[30,40] In the present work, the crystal structure of secondary phases were refined on single-phase α-Ti(Fe,Mn)-D patterns and then it has been fixed during the sequential refinement apart from their abundance. The determined amount of $TiH_2$-type phase variated between 1 and 3 wt.% in all samples, while the oxide-type phase ranges between 1 and 4 wt.%, in good agreement with previous XRD determinations, as reported in **Table 4**.[40] Good agreement is also observed in the cell parameter of the α phase (**Table 4**), confirming the role of manganese in increasing the cell size of the pristine alloy, thus lowering the pressure plateaus of PCI curves.[29]

**Table 2** gathers the results of the Rietveld refinement performed on single-phase patterns collected at ILL in absorption for the γ phase (orthorhombic, *S.G. Cmmm*), during desorption and absorption for the β phase (orthorhombic, *S.G.* $P222_1$) and during desorption for the α phase (cubic, *S.G. Pm*-3*m*). Errors are reported from the Rietveld analysis by using the Berar formula and so considering a multiply factor on standard deviation of 2.75.[44]

Graphical output of Rietveld refinements (including prf files) and the obtained refinement quality factors ($R_b$ for each phase and $R_{wp}$ for the whole pattern) are available at ref.[49].

Absorption and desorption effects on the crystal structure of the $H_2$-carrier material were determined and therefore the values for volume expansion are listed in **Table 2**. Volume expansions do not vary significantly among samples, and they can be defined as 11-15% for the β phase and equal to 17% for the γ phase, with respect to the dehydrogenated α phase.[40] For the β phase, the volume variation is higher on absorption (13-15%) than on desorption (11-12%), a fact that concurs with the higher deuterium content during the absorption step with respect to the desorption one, in agreement with previous findings for the TiFe-D system.[6,18]

Result of Rietveld refinements using ISIS data, which are all multi-phase patterns besides that initial and final states, largely confirmed the results discussed above for data from ILL. It is worth noting that similar anisotropic broadening in the peak shapes of the β phase were observed in the ISIS data.

This confirms that they are due to microstructural features of the phase, rather than instrumental effects. The secondary phases observed in the time-of-flight measurements were also the same. The weight fractions of various phases and corresponding cell parameters are summarized in **Table 4**. They largely coincide with the results obtained with ILL data. Most importantly, the amount of secondary phases is not greater than those reported in **Table 4**, which supports that surface oxidation and/or incomplete activation are the likely cause for incomplete deuterium absorption (see **Figure 3**).

***3.2.3 Structural analysis***

**Figure 4** gathers, for samples A (left) and B (right), the cell volume as well as the total deuterium content ($D_{tot}$/f.u.) and its distribution per site ($D_{site}$/f.u.) for α, β and γ phases, as determined by the Rietveld refinements on patterns collected at ILL, as a function of the scan number. In addition, $TiH_2$-type and $Ti_4Fe_2O$-type phase amounts are also displayed.

The α phase is properly refined in the cubic structure (*S.G. Pm*-3*m*), its cell volume evolution during deuteration and desorption for different Mn contents can be followed comparing **Figure 4-(a)** for sample A, B and **Figure 5-(a)** for sample C, evidencing an increased cell volume when Mn is introduced. **Figure 4-(b)** shows that the deuterium content keeps below 0.10 D/f.u., in good agreement with the literature.

In principle, cell volumes should remain constant during all phase transformations, but the β phase does not observe this rule (**Figure 4-(a)**). During phase transitions on desorption (γ→β→α) the cell volume of the β phase gradually decreases. This variation is less pronounced on increasing the Mn content in the alloy (**Figure 5**). On absorption (α→β→γ), and irrespectively on the Mn content, the β cell volume strongly increases. The total deuterium content and individual contributions from the two different D-sites in the β phase (2*a* and 2*d*) both on absorption and desorption can be observed in **Figure 4-(b)** and **Figure 5-(b)**. Deuterium content varies more significantly over the 2*d* site, especially on absorption. As introduced, different β phases were reported in the literature with two β phases observed on desorption and one on absorption.[18] However the present analysis offers a good description of phase transitions using a unique β phase both in absorption and desorption with orthorhombic structure (*S.G. P*$222_1$). Nonetheless, the occurrence of several β phases with different but close deuterium contents cannot be ruled out, explaining the observed variations of the β cell volume during phase transitions.

The γ phase has been refined in an orthorhombic structure (*S.G. Cmmm*). A good refinement is obtained during cycling, showing almost constant cell parameters and volume for all Mn contents (**Figure 4-(a)** and **Figure 5-(a)**). However, a variation in the total deuterium content can be observed after the second absorption at high scan numbers (**Figure 4-(b)**), which may be connected with either a solid solubilization of deuterium or to residual stress and generated dislocations as discussed in previous neutron studies in ref.[3,12,28].

The γ phase neutron diffraction pattern of sample B ($TiFe_{0.85}Mn_{0.05}$) was also recorded at the high-resolution beamline D2B at ILL (**Figure S3**). Results from D2B are in good agreement with the D1B determination and support the quality of the *in-situ* refinement (**Figure 4-B**, scan number 25). The lower deuterium occupancy obtained from the refinement and visible in scan 25 on **Figure 4-B-(b)** can be justified because of a lower pressure reached in the vial due to a deuterium leak over time between the measurements.

Results on structural evolution highlight that the deuterium content in the γ phase at a given pressure increases with Mn substitution. Higher Mn contents in the alloy allow a higher degree of deuteration, linked to a lower equilibrium pressure, as observed comparing results obtained for sample A ($TiFe_{0.90}$) with those related to sample C ($TiFe_{0.80}Mn_{0.10}$). Hysteresis effects are also observed to diminish upon increasing the Mn content.[40]

In the β phase, a large variation of cell volume during desorption is observed for Mn-free sample A. Upon increasing the Mn content, this variation is almost suppressed. In contrast, on absorption, the cell volume of the β phase gradually increases irrespectively of Mn-content.

Finally, in the α phase, the maximum deuterium content (0.06 D/f.u.) does not change with Mn content, while the cell volume increases with Mn content, due to Fe by Mn substitution.

**Figure 5** shows the phase abundance, deuterium occupancy and cell volume for sample C measured at ISIS. The changes in the phase abundance are markedly different from those observed at ILL, due to the different initial and final states of the sample and to the incomplete hydrogen absorption reaction. On the other hand, the deuterium occupancies and cell volumes provide similar values for various phases, confirming the consistency of the Rietveld refinement results from different data sets. In conclusion, it has been evidenced how the inclusion of Mn enhances the overall capacity and leads to decreased hysteresis compared to the parent TiFe material. This positive substitutional effect is mostly related to the structural modification of the TiFe cell, causing an expansion and enabling easier hydrogen diffusion into the structure to nucleate and form the hydride. A possible mechanism for this behaviour can suppose that both structural and electronic-based effects are induced by Mn-substitution, both by expanding the cell volume and interacting with hydrogen guaranteeing a stronger binding energy, thus lowering the enthalpy and the equilibrium pressure at which hydrogen can be absorbed. Since lower hydrogenation pressures are required after Mn substitution, higher capacities can be reached by forming the di-hydride (full hydrogenation) at lower pressure. Similarly, the effect of Mn substitution on hysteresis can be supposed to be connected with its chemistry and effect on mechanical properties of the alloy, guaranteeing a reversible hydrogenation in a narrower pressure range between absorption and desorption. Especially, it can be related to an effect of Mn in reducing plastic deformation during hydrogen absorption.

## 4. Conclusions

Reversible deuterium sorption and phase transformations have been successfully characterized in three different TiFe-type alloys by *in situ* neutron diffraction, while recording the Pressure-Composition Isotherms at RT.

Neutron diffraction analysis demonstrates that different contents of Mn do not significantly change the volumetric expansion during phase transitions. Mn substitution affects the deuterium content for the γ phase and the cell evolution for the β phase. This is linked with the lowering of the high-plateau pressure with Mn content.

Analyses of Mn content on structural and hydrogenation properties proved that, increasing the Mn content, a higher deuterium capacity can be obtained, while structural features are mostly retained. Especially regarding the β phase, Mn substitution favors a stable deuterium occupancy over the two available 2*a* and 2*d* sites, compared to Mn-free sample where a large variation in deuterium occupancy is evidenced over 2*d* site, particularly during desorption.

The study confirms that the deuterated structures of the γ phase upon absorption, β and α phase upon desorption, correspond to S.G. *Cmmm*, $P222_1$ and *Pm*-3*m*, respectively.

In addition to the determination of structural and composition evolution, the fundamental knowledge of hydrogenation and dehydrogenation mechanisms of $TiFe_{0.90-x}Mn_x$ alloys is relevant for the scientific community and helps in the final design of the material, since the determined volume expansion and reduction of the alloy during $H_2$ uptake and release must be considered during tank design.

## Acknowledgements

HyCARE project has received funding from the Fuel Cells and Hydrogen 2 Joint Undertaking (now Clean Hydrogen Partnership) under Grant Agreement No 826352. This Joint Undertaking receives support from the European Union's Horizon 2020 Research and Innovation program, Hydrogen Europe and Hydrogen Europe Research.

The authors wish to thank Laetitia Laversenne at ILL, and Ivan da Silva Gonzalez and Mark Kibble at ISIS for their help and support during beamtimes and *in-situ* experimental setup.

This manuscript honours the memory of our colleague Michel Latroche, who significantly contributed to the work presented herein.

## Electronic Supporting Information Description

ESI includes the comparison of PCI determined capacity at ILL and total deuterium content determined by Rietveld refinement as a function of scan number for sample A, B, and C; zoomed region of the Rietveld refinement to evidence asymmetric peak profiles; and the neutron diffraction pattern and Rietveld refinement of the γ phase of sample B recorded at ILL beamline D2B at RT.

## Supplementary data

Supplementary data to this article can be found online at https://dx.doi.org/10.5291/ILL-DATA.5-22-771 and https://doi.org/10.5286/ISIS.E.RB1920559, which contains raw data recorded at ILL and ISIS respectively.[41,45]

Origin resume file with all determination and graphs, and Rietveld refinements outputs are available at https://doi.org/10.5281/zenodo.6620531.[49] From the cited dataset, Rietveld refinements graphs for all patterns analysed can be downloaded, ".out" refinement files can also be found and downloaded for further refinement details and related $R_p$, $R_{exp}$, GoF factors of each pattern which are consistent with the ones reported of pure phases.

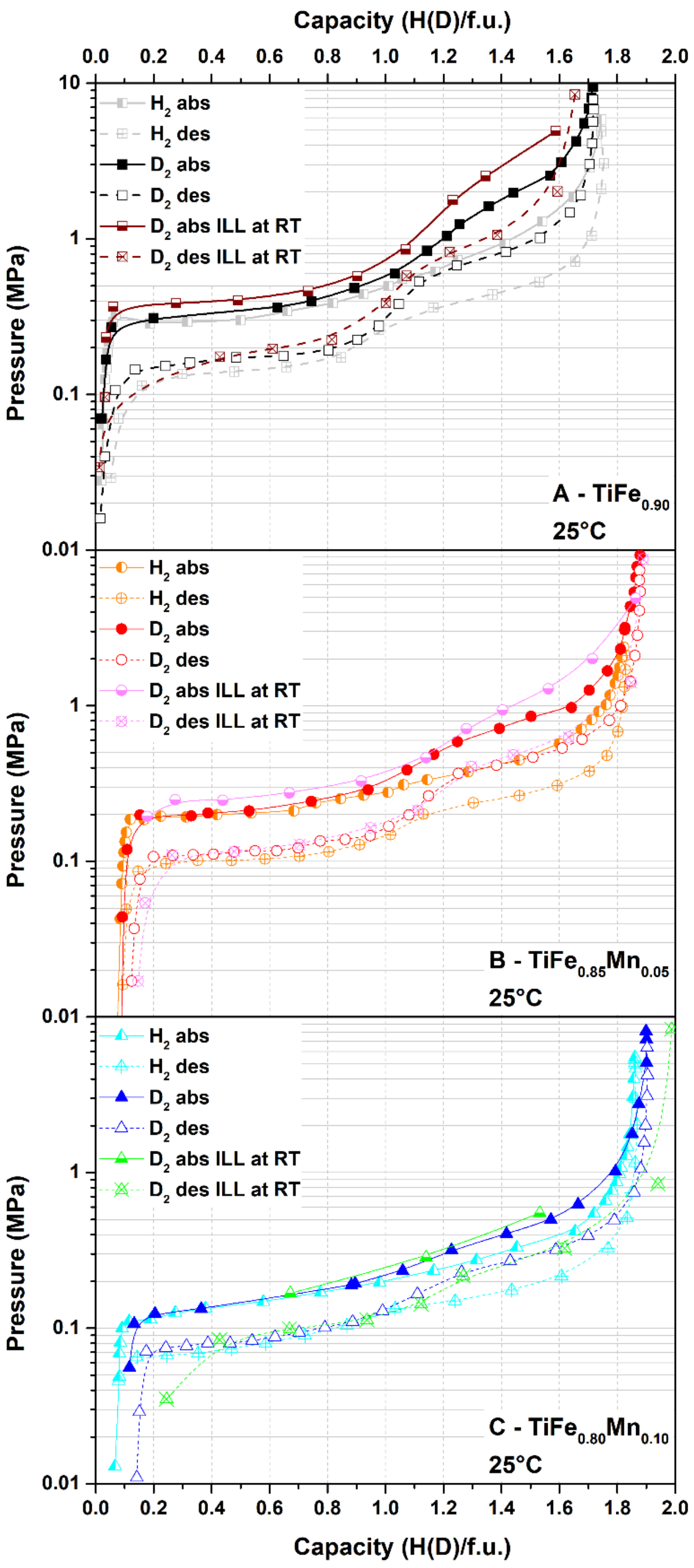

**Figure 1** – Absorption (abs, full and half-full symbols) and desorption (des, empty and crossed symbols) PCI curves at 25 °C for samples A, B and C during in-lab deuteration ($D_2$, full/empty points), *in-situ* deuteration at ILL at RT ($D_2$, bottom-half-full/crossed-x points) and in-lab hydrogenation at 25 °C ($H_2$, left-half-full and crossed-+ points, from ref.[40] and datasets from [48]). Solid lines in absorption and dashed lines in desorption are a guide for the reader, errors are within data points.

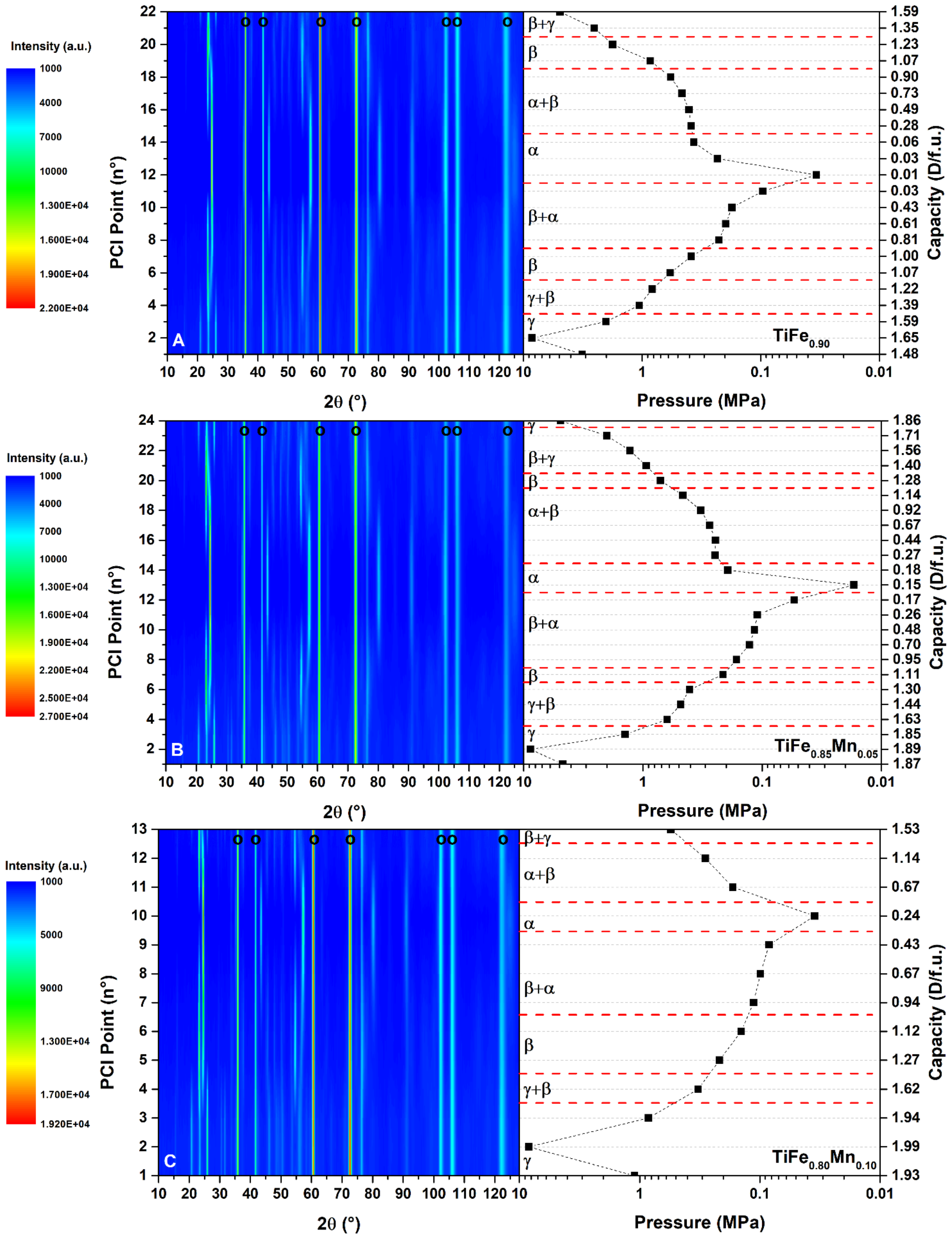

**Figure 2** – *In-situ* neutron diffraction patterns (counter plot, λ=1.286 Å) and PCI curve at RT during deuterium absorption/desorption at ILL between 0.02 and 9 MPa for sample A (top), B (middle) and C (bottom) at RT. Diffraction lines from the stainless-steel sample holder are marked by a circle (o), errors are within data points.

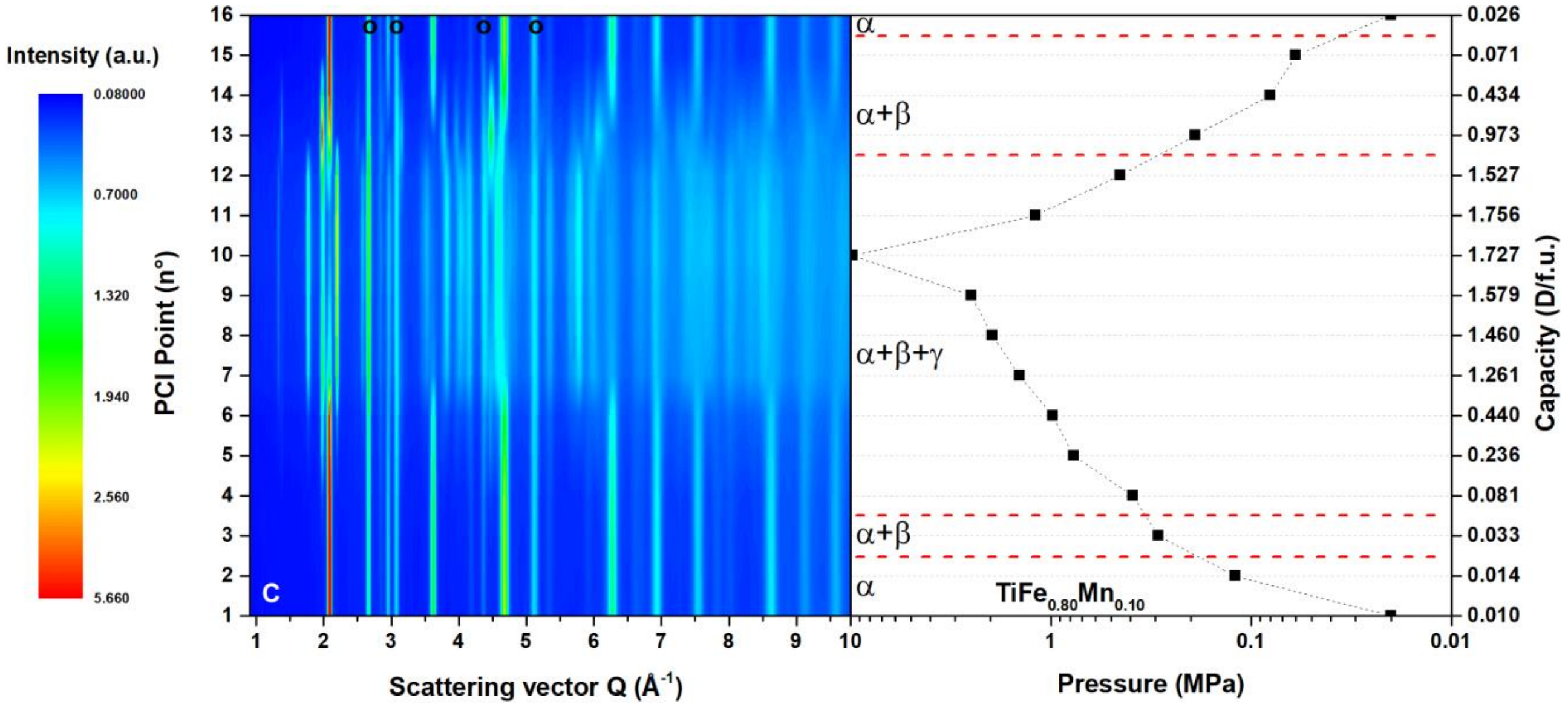

**Figure 3** – *In-situ* neutron diffraction patterns (counter plot) and PCI curve during deuterium absorption/desorption at ISIS between 0.02 and 9.8 MPa for sample C at RT. Diffraction lines from the aluminium sample holder are marked by a circle (o), errors are within data points.

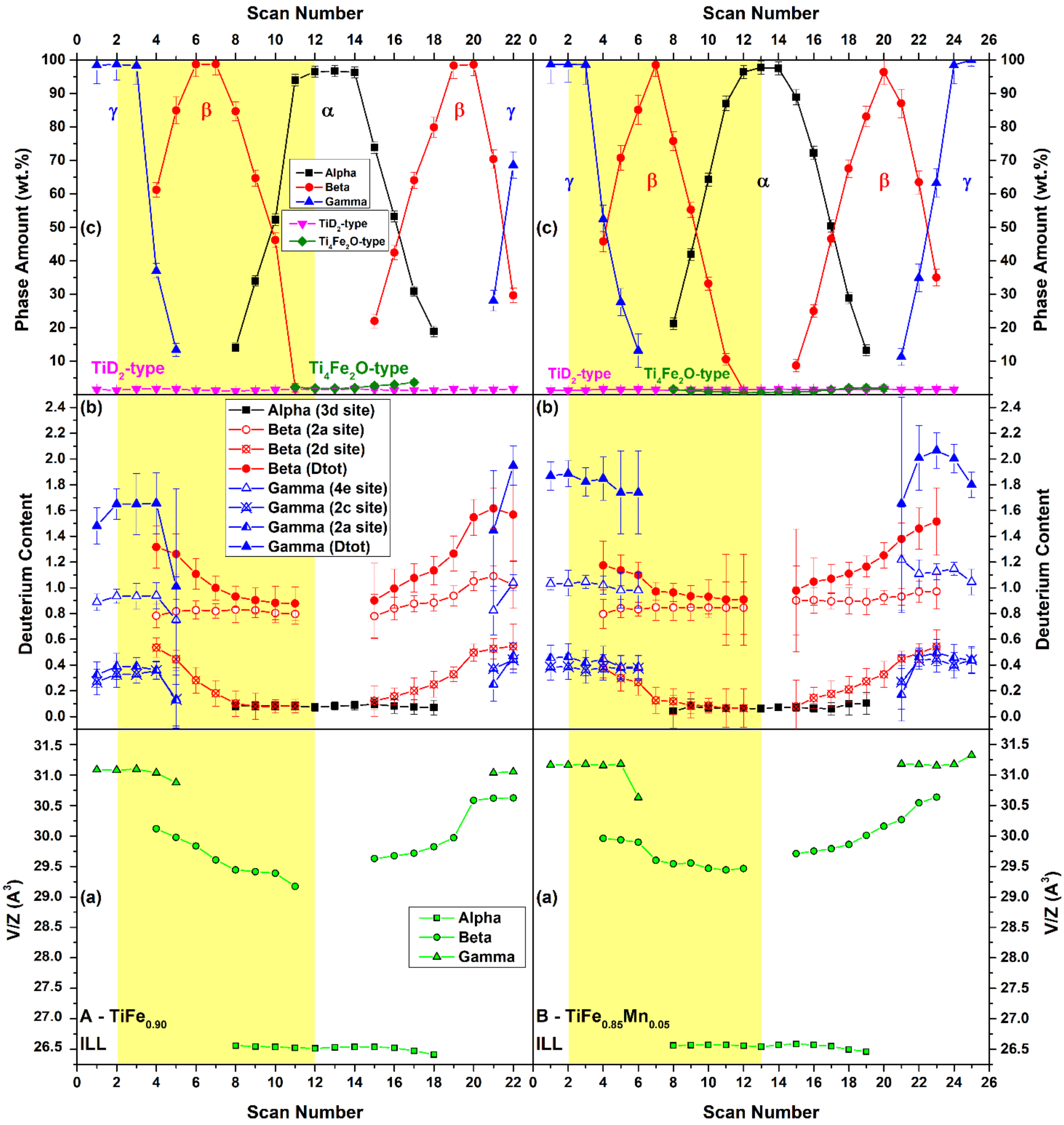

**Figure 4** – Rietveld refinement results as a function of the scan number of the *in-situ* neutron diffraction analysis at ILL for sample A (left) and B (right): (a) cell volume for phase α , β and γ, errors are within data points, (b) total and site deuterium content in phase α, β and γ, (c) α, β, γ, $TiD_2$-type and $Ti_4Fe_2O$-type phase amount. Yellow area is evidencing desorption determinations, while white area refers to absorption. Scan numbers correspond to PCI points reported in the manuscript.

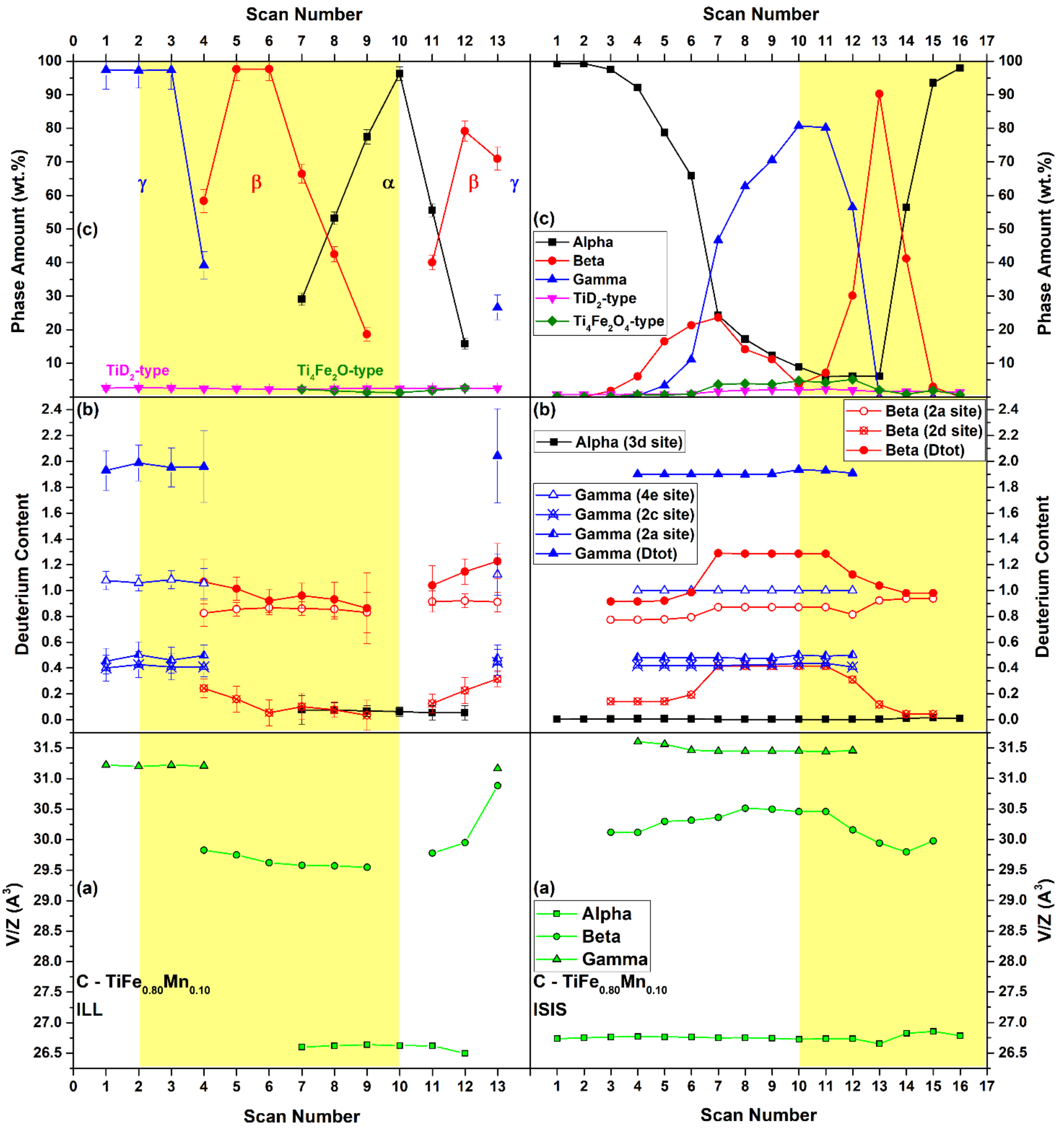

**Figure 5** – Rietveld refinement results as a function of the scan number of the *in-situ* neutron diffraction analysis at ILL (left) and ISIS (right) for sample C: (a) cell volume for phase α, β and γ, (b) total and site deuterium content in phase α, β and γ, (c) α, β, γ, $TiD_2$-type and $Ti_4Fe_2O$-type phase amount. Errors for cell volume and ISIS determinations are within data points. Yellow area is evidencing desorption determinations, while white area refers to absorption. Scan numbers correspond to PCI points reported in the manuscript.

**Table 1** – Literature review of TiFe and its hydrides structures and structural parameters.

| Structure | *S.G.* | *a* (Å) | *b* (Å) | *c* (Å) | *β/γ* (°) | Atomic positions | Formula | Ref. |
|---|---|---|---|---|---|---|---|---|
| **α phase** | | | | | | | | |
| **Cubic** | *Pm*-3*m* | 2.959 | | | | | TiFe | [11] |
| **Cubic** | *Pm*-3*m* | 2.975 | | | | | TiFe | [10] |
| **Cubic** | *Pm*-3*m* | 2.976 | | | | | TiFe | [12–14] |
| **Cubic** | *Pm*-3*m* | 2.979 | | | | | TiFe | [15] |
| **Cubic** | *Pm*-3*m* | 3.011 | | | | | TiFe | [16] |
| **Cubic** | *Pm*-3*m* | 2.978 | | | | | $TiFeH_{0.02}$ | [52] |
| **Cubic** | *Pm*-3*m* | 2.977 | | | | | $TiFeH_{0.04}$ | [14] |
| **Cubic** | *Pm*-3*m* | 2.980 | | | | | $TiFeH_{0.057}$ | [17,18] |
| **Cubic** | *Pm*-3*m* | 2.979 | | | | | $TiFeH_{0.06}$ | [22] |
| **Cubic** | *Pm*-3*m* | 2.985 | | | | | $TiFe_{0.885}Mn_{0.115}$ | [19] |
| **Cubic** | *Pm*-3*m* | 2.986 | | | | | $TiFe_{0.80}Mn_{0.20}$ | [31] |
| **Cubic** | *Pm*-3*m* | 2.983 | | | | | $TiFe_{0.94}Ce_{0.06}H_{0.03}$ | [52] |
| **β phase** | | | | | | | | |
| **Tetragonal** | | 3.180 | | 8.730 | | | $TiFeH_{1.06}$ | [6] |
| **Orthorhombic** | | 4.500 | 4.300 | 2.980 | | | $TiFeH_{0.67-1.02}$ | [18] |
| **Orthorhombic** | *Pmcm* | 2.954 | 4.538 | 4.381 | | | $TiFeH_{0.94}$ | [22] |
| **Orthorhombic** | *Pmcm* | 2.991 | 4.556 | 4.398 | | | $Ti_{1.04}Fe_{0.96}H_{1.15}$ | [23] |
| **Orthorhombic** | *Pmcm* | 3.094 | 4.513 | 4.391 | | | $TiFeH_{1.4}$ | [22] |
| **Orthorhombic** | *Pmcm* | 2.968 | 4.535 | 4.382 | | | $Ti_{0.98}Fe_{1.02}H_{1.03}$ | [23] |
| **Orthorhombic** | *Pmcm* | 2.973 | 4.561 | 4.407 | | | $Ti_{1.02}Fe_{0.98}H_{1.03}$ | [23] |
| **Orthorhombic** | $P222_1$<br>$Pmc2_1$<br>*Pmcm*<br>*P*2*cm* | 2.956 | 4.543 | 4.388 | | $y_{Fe, 2c}$ = 0.206<br>$x_{D1, 2a}$ = 0.058<br>$y_{D1, 2a}$ = 0.537 | $TiFeH_{1.0\pm0.04}$ | [20] |
| **Orthorhombic** | $P222_1$<br>$Pmc2_1$<br>*P*2/*c* | 2.966 | 4.522 | 4.370 | | $y_{Ti, 2d}$ = 0.757<br>$y_{Fe, 2c}$ = 0.294<br>$y_{D2, 2d}$ = 0.300 | $TiFeH_{1.0}$ | [10,12,21] |
| **Orthorhombic** | $P222_1$ | 2.978 | 4.549 | 4.417 | | | $TiFeH_{1.0}$ | [16] |
| **Orthorhombic** | $P222_1$ | 2.967 | 4.522 | 4.371 | | $y_{Ti, 2d}$ = 0.757<br>$y_{Fe, 2c}$ = 0.294 | $TiFeH_{1.0}$ | [12] |
| **Orthorhombic** | $P222_1$ | 2.932 | 4.515 | 4.276 | | | $TiFeH_{1.0}$ | [11] |
| **Orthorhombic** | $P222_1$ | 3.088 | 4.515 | 4.391 | | $y_{Ti, 2d}$ = 0.750<br>$y_{Fe, 2c}$ = 0.303<br>$y_{D2, 2d}$ = 0.297 | $TiFeH_{1.37}$ | [21] |
| **Orthorhombic** | $Pb2_1m$<br>$P22_12$ | 4.475 | 4.319 | 2.925 | | | $TiFeH_{1.0}$ | [13] |
| **Orthorhombic** | $Pb2_1m$<br>$P22_12$ | 4.543 | 4.369 | 2.972 | | | $TiFeH_{1.0}$ | [13] |
| **Orthorhombic** | $Pmc2_1$<br>*P*2*cm* | 2.995 | 4.531 | 4.397 | | $y_{Ti, 2d}$ = 0.249<br>$y_{Fe,Mn, 2c}$ = 0.694<br>$x_{D1, 2a}$ = 0.037<br>$y_{D2, 2d}$ = 0.555<br>$y_{D2, 2d}$ = 0.737 | $TiFe_{0.90}Mn_{0.10}H_{1.25}$ | [19] |
| **γ phase** | | | | | | | | |
| **Cubic** | | 6.610 | | | | | $TiFeH_{1.93}$ | [6] |
| **Orthorhombic** | | 7.000 | 6.200 | 3.100 | | | $TiFeH_{1.38-1.88}$ | [18] |
| **Orthorhombic** | *Cmmm* | 7.043 | 6.237 | 2.832 | | $x_{Ti,4h}$ = 0.284<br>$y_{Fe,4i}$ = 0.211 | $TiFeH_{1.9}$ | [15] |
| **Orthorhombic** | *Cmmm* | 7.029 | 6.233 | 2.835 | | $x_{Ti,4h}$ = 0.223<br>$y_{Fe,4i}$ = 0.2887 | $TiFeH_{1.94}$ | [24,25] |
| **Orthorhombic** | *Cmmm* | 7.123 | 6.302 | 2.819 | | | $TiFeH_{2.0}$ | [16] |
| **Orthorhombic** | *Cmmm* | 6.980 | 6.115 | 2.813 | | | $TiFeH_{2.0}$ | [11] |
| **Monoclinic** | *P2/m* | 4.706 | 2.835 | 4.697 | | $x_{Ti,2n}$ = 0.278<br>$z_{Ti,2n}$ = 0.217<br>$x_{Fe,2m}$ = 0.199<br>$z_{Fe,2m}$ = 0.727 | $TiFeH_{1.73}$ | [21] |
| **Monoclinic** | *P2/m* | 4.704 | 2.831 | 4.704 | 96.94 | $y_{Ti,2n}$ = 0.283<br>$z_{Ti,2n}$ = 0.217<br>$x_{Fe,2m}$ = 0.222<br>$z_{Fe,2m}$ = 0.701 | $TiFeH_{1.9}$ | [15] |
| **Monoclinic** | *P2/m* | 4.713 | 2.834 | 4.713 | 97.12 | | $TiFeH_{1.9}$ | [22] |
| **Monoclinic** | *P2/m* | 4.704 | 2.830 | 4.704 | 96.98 | $y_{Ti,2n}$ = 0.294<br>$z_{Ti,2n}$ = 0.229<br>$x_{Fe,2m}$ = 0.203<br>$z_{Fe,2m}$ = 0.721 | $TiFeH_{1.9\pm0.01}$ | [26] |
| **Monoclinic** | *P2/m* | 4.712 | 2.841 | 4.685 | 96.88 | $x_{Ti,2n}$ = 0.28<br>$z_{Ti,2n}$ = 0.22<br>$x_{Fe,2m}$ = 0.214<br>$z_{Fe,2m}$ = 0.712 | $TiFeH_{1.94}$ | [24] |
| **Monoclinic** | *P2/m* | 4.733 | 2.843 | 4.688 | 97.62 | | $Ti_{1.04}Fe_{0.96}H_{1.98}$ | [23] |
| **Monoclinic** | *P2/m* | 4.708 | 4.697 | 2.835 | 97.05 | $x_{Ti,2n}$ = 0.274<br>$z_{Ti,2n}$ = 0.231<br>$x_{Fe,2m}$ = 0.200<br>$z_{Fe,2m}$ = 0.736 | $TiFeH_{2.0}$ | [13,27] |
| **Monoclinic** | *P2/m* | 4.714 | 2.837 | 4.714 | 97.15 | $x_{Ti,2n}$ = 0.267<br>$z_{Ti,2n}$ = 0.240<br>$x_{Fe,Mn,2m}$ = 0.181<br>$z_{Fe,Mn,2m}$ = 0.744 | $TiFe_{0.90}Mn_{0.10}H_{1.72}$ | [19] |
| **Monoclinic** | *P2/m* | 4.702 | 2.834 | 4.719 | 97.31 | | $TiFe_{0.80}Mn_{0.20}H_{1.8}$ | [31] |
| **Monoclinic** | *P2/m* | 4.713 | 2.834 | 4.717 | 97.11 | | $TiFe_{0.79}Mn_{0.15}H_{2.0}$ | [23] |

**Table 2** – Structural details of single-phase Ti(Fe,Mn)-D phases for samples A, B, and C as determined in this work from *in-situ* neutron diffraction during deuterium absorption and desorption (at ILL and ISIS). Errors in the last digit are given between brackets. Complementary structural data are reported in **Table S1**.

| Sample | *Pressure (MPa)* | $D_{tot}$ (D/f.u.) | *a* (Å) | *b* (Å) | *c* (Å) | *V/Z* (Å$^3$) | Volume expansion (%) | Refined atomic positions | Rietveld R-factors |
|---|---|---|---|---|---|---|---|---|---|
| ***γ (1st absorption, S.G. Cmmm)*** | | | | | | | | | |
| **A) $TiFe_{0.90}$** | 8.47 | 1.7(1) | 7.036(1) | 6.238(1) | 2.833(1) | 31.081(3) | 17.2 | $y_{Ti,4h}$ = 0.211(2)<br>$y_{Ti/Fe,4i}$ = 0.285(1) | $R_B$ = 9.82<br>$R_{wp}$ = 11.1 |
| **B) $TiFe_{0.85}Mn_{0.05}$** | 8.77 | 1.9(1) | 7.049(3) | 6.239(3) | 2.834(1) | 31.164(7) | 17.4 | $y_{Ti,4h}$ = 0.213(2)<br>$y_{Ti/Fe/Mn,4i}$ = 0.286(1) | $R_B$ = 11.3<br>$R_{wp}$ = 12.7 |
| **C) $TiFe_{0.80}Mn_{0.10}$** | 8.37 | 2.0(1) | 7.056(5) | 6.238(4) | 2.835(1) | 31.196(9) | 17.2 | $y_{Ti,4h}$ = 0.212(2)<br>$y_{Ti/Fe/Mn,4i}$ = 0.285(1) | $R_B$ = 13.0<br>$R_{wp}$ = 14.8 |
| ***β (desorption, S.G. P2221)*** | | | | | | | | | |
| **A) $TiFe_{0.90}$** | 0.39 | 1.0(1) | 2.982(1) | 4.524(4) | 4.388(4) | 29.608(9) | 11.7 | $y_{Ti,2d}$ = 0.747(3)<br>$y_{Ti/Fe,2c}$ = 0.291(1) | $R_B$ = 11.3<br>$R_{wp}$ = 12.9 |
| **B) $TiFe_{0.85}Mn_{0.05}$** | 0.21 | 1.0(1) | 2.978(1) | 4.524(3) | 4.394(3) | 29.602(7) | 11.5 | $y_{Ti,2d}$ = 0.749(3)<br>$y_{Ti/Fe/Mn,2c}$ = 0.296(1) | $R_B$ = 11.7<br>$R_{wp}$ = 13.7 |
| **C) $TiFe_{0.80}Mn_{0.10}$** | 0.14 | 0.9(1) | 2.979(1) | 4.521(4) | 4.398(4) | 29.619(9) | 11.2 | $y_{Ti,2d}$ = 0.745(3)<br>$y_{Ti/Fe/Mn,2c}$ = 0.293(1) | $R_B$ = 13.5<br>$R_{wp}$ = 15.6 |
| ***α (desorption, S.G. Pm-3m)*** | | | | | | | | | |
| **A) $TiFe_{0.90}$ – ILL** | 0.03 | 0.07(3) | 2.982(1) | | | 26.513(1) | | | $R_B$ = 8.39<br>$R_{wp}$ = 10.0 |
| **A) $TiFe_{0.90}$ – ISIS** | | 0.001(1) | 2.984(1) | | | 26.589(1) | | | $R_p$ = 4.9<br>$R_{wp}$ = 7.3 |
| **B) $TiFe_{0.85}Mn_{0.05}$ – ILL** | 0.02 | 0.06(2) | 2.983(1) | | | 26.543(1) | | | $R_B$ = 8.55<br>$R_{wp}$ = 10.3 |
| **B) $TiFe_{0.85}Mn_{0.05}$ – ISIS** | | 0.000(1) | 2.988(1) | | | 26.664(1) | | | $R_p$ = 3.6<br>$R_{wp}$ = 5.3 |
| **C) $TiFe_{0.80}Mn_{0.10}$ – ILL** | 0.03 | 0.06(4) | 2.986(1) | | | 26.627(1) | | | $R_B$ = 10.4<br>$R_{wp}$ = 12.6 |
| **C) $TiFe_{0.80}Mn_{0.10}$ – ISIS** | | 0.003(1) | 2.990(1) | | | 26.737(1) | | | $R_p$ = 4.9<br>$R_{wp}$ = 7.3 |
| ***β (absorption, S.G. P2221)*** | | | | | | | | | |
| **A) $TiFe_{0.90}$** | 1.78 | 1.5(1) | 3.093(2) | 4.507(4) | 4.388(3) | 30.586(9) | 15.4 | $y_{Ti,2d}$ = 0.747(3)<br>$y_{Ti/Fe,2c}$ = 0.291(1) | $R_B$ = 13.0<br>$R_{wp}$ = 14.1 |
| **B) $TiFe_{0.85}Mn_{0.05}$** | 0.71 | 1.3(1) | 3.039(1) | 4.512(4) | 4.400(3) | 30.162(9) | 13.6 | $y_{Ti,2d}$ = 0.749(3)<br>$y_{Ti/Fe/Mn,2c}$ = 0.296(1) | $R_B$ = 14.0<br>$R_{wp}$ = 16.2 |
| ***γ (2nd absorption, S.G. Cmmm)*** | | | | | | | | | |
| **B) $TiFe_{0.85}Mn_{0.05}$ (D1B)** | 4.90 | 2.0(1) | 7.048(4) | 6.242(3) | 2.834(1) | 31.174(8) | 17.4 | $y_{Ti,4h}$ = 0.213(1)<br>$y_{Ti/Fe/Mn,4i}$ = 0.286(1) | $R_B$ = 12.4<br>$R_{wp}$ = 14.2 |
| **B) $TiFe_{0.85}Mn_{0.05}$ (D2B)** | | 1.8(1) | 7.064(1) | 6.247(1) | 2.840(1) | 31.328(3) | 18.0 | $y_{Ti,4h}$ = 0.220(1)<br>$y_{Ti/Fe/Mn,4i}$ = 0.285(1) | $R_B$ = 14.9<br>$R_{wp}$ = 15.1 |

**Table 3** – Site occupancy factors (SOF) for metal elements in TiFe and $TiH_2$ phases used in Rietveld refinements (based on EPMA determination of ref.[40]).

| Sample | | Phase | Site | $SOF_{Fe}$ | $SOF_{Ti}$ | $SOF_{Mn}$ |
|---|---|---|---|---|---|---|
| **A (s2[40])** | **$TiFe_{0.90}$** | TiFe | 1*a** | 0.964 | 0.036 | 0 |
| | | | 1*b** | 0.000 | 1.000 | 0 |
| | | $TiH_2$ | 4*a* | 0.200 | 0.800 | 0 |
| **B (s6[40])** | **$TiFe_{0.85}Mn_{0.05}$** | TiFe | 1*a** | 0.912 | 0.038 | 0.050 |
| | | | 1*b** | 0.000 | 1.000 | 0 |
| | | $TiH_2$ | 4*a* | 0.190 | 0.787 | 0.023 |
| **C (s9[40])** | **$TiFe_{0.80}Mn_{0.10}$** | TiFe | 1*a** | 0.877 | 0.034 | 0.090 |
| | | | 1*b** | 0.000 | 1.000 | 0 |
| | | $TiH_2$ | 4*a* | 0.170 | 0.780 | 0.050 |

*Wyckoff sites 1*a* and 1*b* refer to the α phase (S.G. *Pm-3m*). Same SOF values have been used for related 2*c* and 2*d* sites for the β phase (S.G. $P222_1$) and 4h and 4i sites for the γ phase (S.G. *Cmmm*)

**Table 4** – Secondary phase amount and lattice parameters as determined from neutron diffraction (ND) at ILL and ISIS reported in this work. X-ray diffraction (XRD) analysis performed in ref.[40,48,53] are reported for comparison. Range of values are reported (minimum and maximum values) when multiple *in-situ* determinations were performed.

| Sample | | ND TiFe *a* (Å) | XRD TiFe *a* (Å) | ND $TiD_2$ *a* (Å) | ND $TiD_2$ wt.% | XRD $TiH_2$ wt.% | ND $Ti_4Fe_2O$ *a* (Å) | ND $Ti_4Fe_2O$ wt.% | XRD $Ti_4Fe_2O$ wt.% |
|---|---|---|---|---|---|---|---|---|---|
| **A – ILL** | **$TiFe_{0.90}$** | 2.978-2.984 | 2.982(7) | 4.407 | 1.1-1.8 | 5.0±0.5 | 11.841 | 1.8-3.7 | 4.0±0.5 |
| **A – ISIS** | **$TiFe_{0.90}$** | 2.985(1) | | 4.377 | 0.6 | | 11.924 | 0.9 | |
| **B – ILL** | **$TiFe_{0.85}Mn_{0.05}$** | 2.980-2.985 | 2.985(6) | 4.402 | 1.3-1.7 | 2.8±0.5 | 11.855 | 0.7-2.1 | 2.4±0.5 |
| **B - ISIS** | **$TiFe_{0.85}Mn_{0.05}$** | 2.988(1) | | 4.393 | 0.1 | | 11.780 | 0.3 | |
| **C - ILL** | **$TiFe_{0.80}Mn_{0.10}$** | 2.981-2.987 | 2.987(0) | 4.372 | 2.3-2.7 | 2.8±0.5 | 11.865 | 1.3-2.6 | 1.7±0.5 |
| **C - ISIS** | **$TiFe_{0.80}Mn_{0.10}$** | 2.987-2.994 | | 4.389-4.415 | 0.7-2.3 | | 11.574-11.928 | 0.05-3.03 | |

# ESI

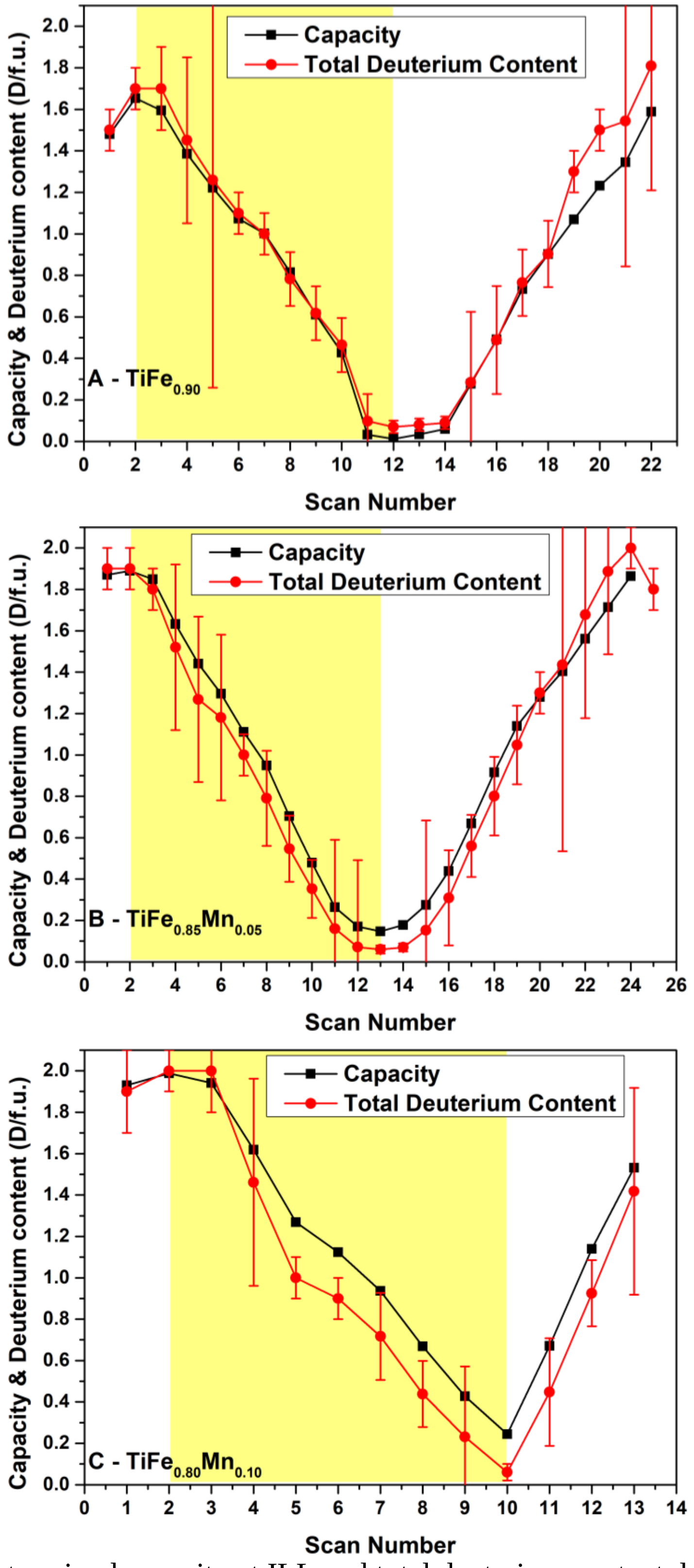

**Figure S1** – PCI determined capacity at ILL and total deuterium content determined by Rietveld refinement as a function of scan number for sample A, B, and C. Yellow area is evidencing desorption determinations, while white area refers to absorption. Scan numbers correspond to PCI points reported in the manuscript.

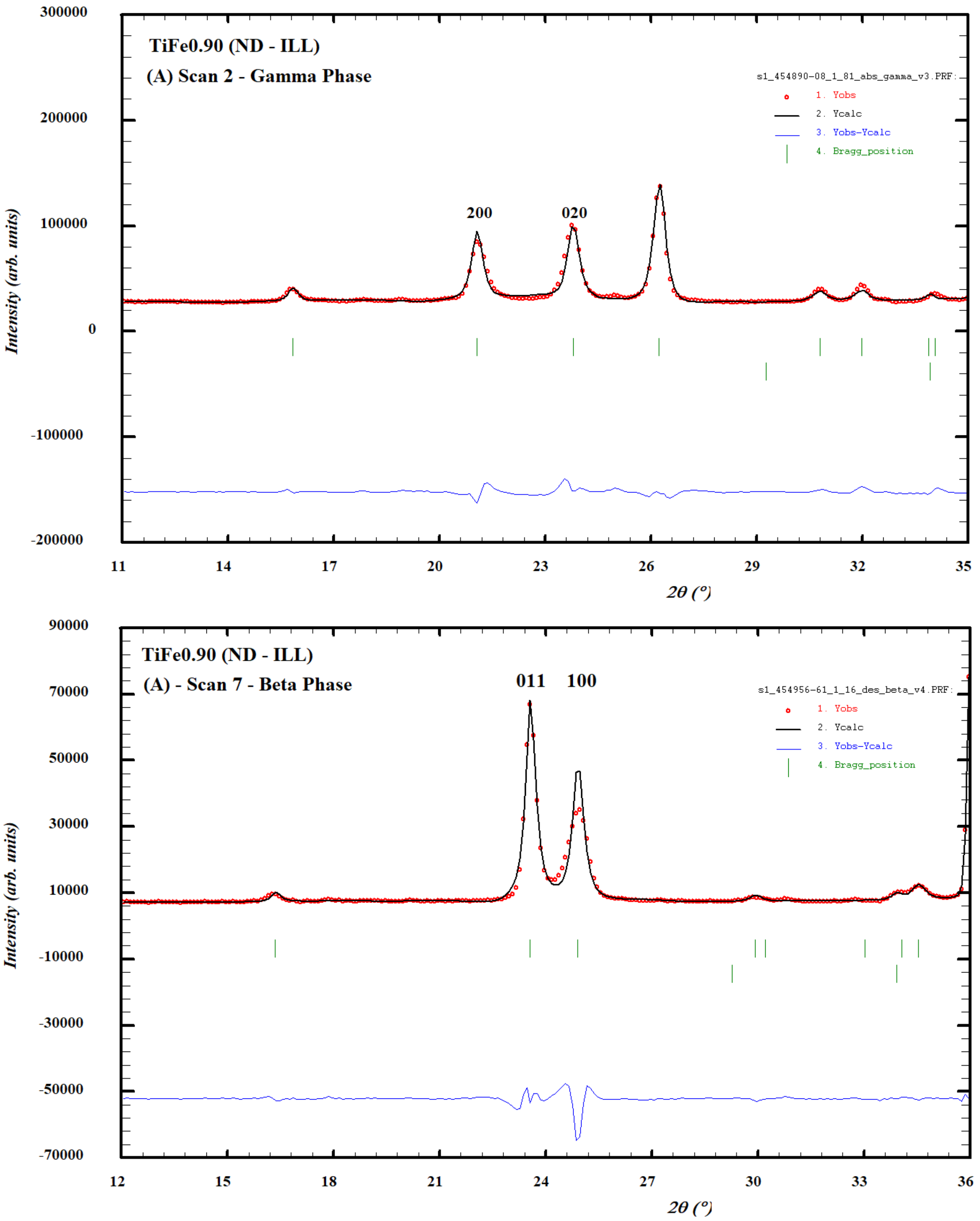

**Figure S2** – Zoomed region of the Rietveld refinement on β (down) and γ (up) phase from scan 7 and 2, respectively. Asymmetric peak profiles can be appreciated.

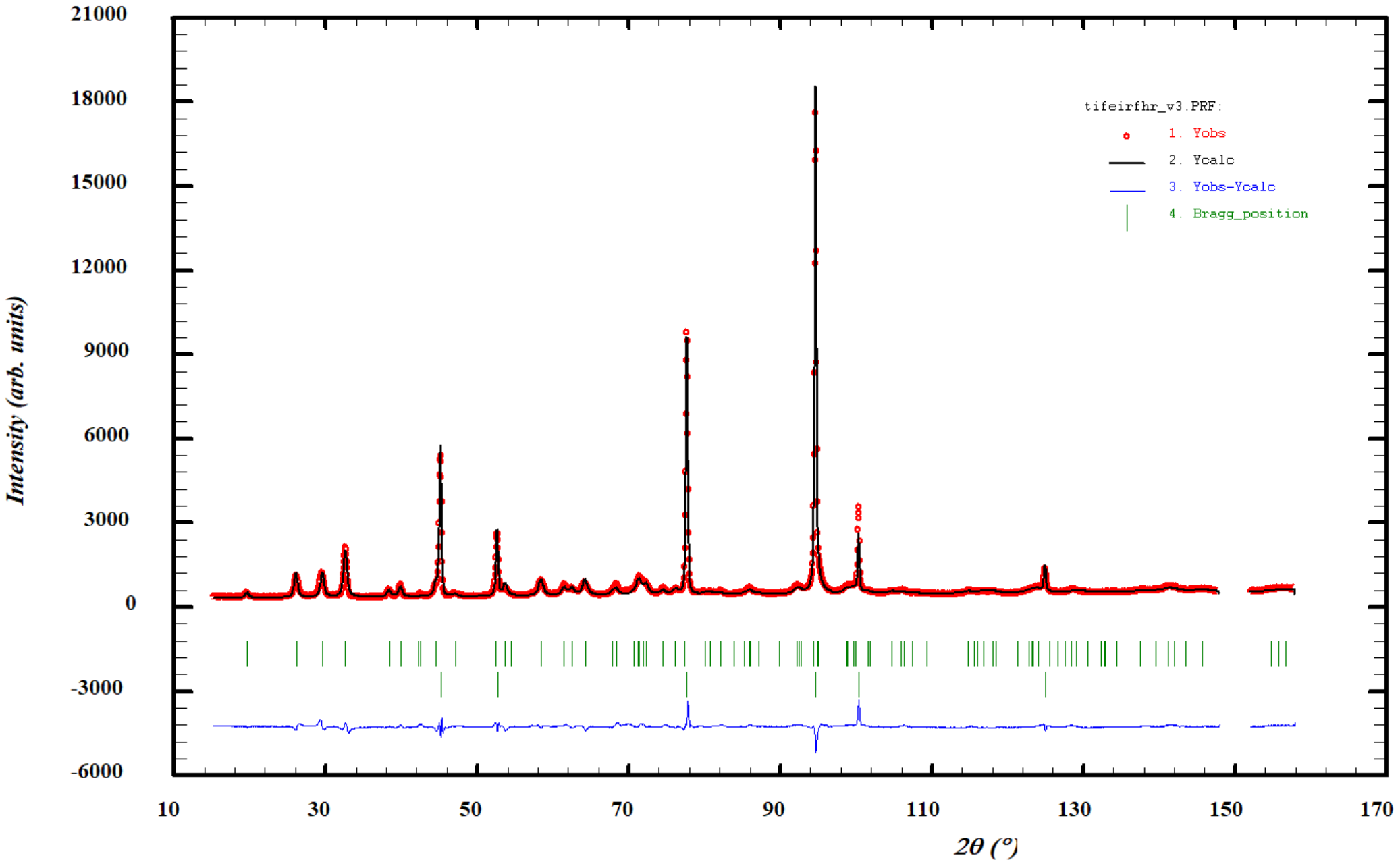

**Figure S3** – Neutron diffraction pattern and Rietveld refinement of the γ phase of sample B (scan number 25, $TiFe_{0.85}Mn_{0.05}$) recorded at ILL beamline D2B at RT.

**Table S1** – Structural details (atom coordinates, thermal factor, occupancies) of single-phase Ti(Fe,Mn)-D phases for samples A, B, and C as determined in this work from *in-situ* neutron diffraction during deuterium absorption and desorption (at ILL and ISIS). Errors in the last digit are given between brackets.

## *γ (1st absorption, S.G. Cmmm)*

### A) $TiFe_{0.90}$

| Atom | X | Y | Z | Biso | Occ | Capacity from PCI [Pressure (MPa)] |
|---|---|---|---|---|---|---|
| *Ti4h* | 0.211(2) | 0 | 0.5 | 0.1 | 4.000 | |
| *Fe4i* | 0 | 0.285(1) | 0 | 0.1 | 3.856 | |
| *Ti4i* | 0 | 0.285(1) | 0 | 0.1 | 0.144 | |
| *D14e* | 0.25 | 0.25 | 0 | 2.0 | 3.7(1) | |
| *D12c* | 0.5 | 0 | 0.5 | 2.0 | 1.3(1) | |
| *D12a* | 0 | 0 | 0 | 2.0 | 1.6(1) | |
| *Dtotal/4* | | | | | 1.7(1) | 1.7(1) [8.47] |

### B) $TiFe_{0.85}Mn_{0.05}$

| Atom | X | Y | Z | Biso | Occ | Capacity from PCI [Pressure (MPa)] |
|---|---|---|---|---|---|---|
| *Ti4h* | 0.213(1) | 0 | 0.5 | 0.4 | 4.000 | |
| *Fe4i* | 0 | 0.286(1) | 0 | 0.4 | 3.648 | |
| *Ti4i* | 0 | 0.286(1) | 0 | 0.4 | 0.152 | |
| *Mn4i* | 0 | 0.286(1) | 0 | 0.4 | 0.200 | |
| *D14e* | 0.25 | 0.25 | 0 | 2.0 | 4.1(1) | |
| *D12c* | 0.5 | 0 | 0.5 | 2.0 | 1.6(1) | |
| *D12a* | 0 | 0 | 0 | 2.0 | 1.9(1) | |
| *Dtotal/4* | | | | | 1.9(1) | 1.9(1) [8.77] |

### C) $TiFe_{0.80}Mn_{0.10}$

| Atom | X | Y | Z | Biso | Occ | Capacity from PCI [Pressure (MPa)] |
|---|---|---|---|---|---|---|
| *Ti4h* | 0.213(1) | 0 | 0.5 | 0.1 | 4.000 | |
| *Fe4i* | 0 | 0.285(1) | 0 | 0.1 | 3.508 | |
| *Ti4i* | 0 | 0.285(1) | 0 | 0.1 | 0.136 | |
| *Mn4i* | 0 | 0.285(1) | 0 | 0.1 | 0.360 | |
| *D14e* | 0.25 | 0.25 | 0 | 2 | 4.2(1) | |
| *D12c* | 0.5 | 0 | 0.5 | 2 | 1.7(1) | |
| *D12a* | 0 | 0 | 0 | 2 | 2.0(1) | |
| *Dtotal/4* | | | | | 2.0(1) | 2.0(1) [8.37] |

## ***β (desorption, S.G. P222$_1$)***

**A) TiFe$_{0.90}$**

| Atom | X | Y | Z | Biso | Occ | Capacity from PCI [Pressure (MPa)] |
|---|---|---|---|---|---|---|
| *Ti2d* | 0.5 | 0.747(3) | 0.25 | 0.1 | 2.000 | |
| *Fe2c* | 0 | 0.291(1) | 0.25 | 0.1 | 1.928 | |
| *Ti2c* | 0 | 0.291(1) | 0.25 | 0.1 | 0.072 | |
| *D12a* | 0 | 0 | 0 | 1.8 | 1.6(1) | |
| *D22d* | 0.5 | 0.3 | 0.25 | 1.8 | 0.4(1) | |
| *Dtotal/2* | | | | | 1.0(1) | 1.0(1) [0.39] |

**B) TiFe$_{0.85}$Mn$_{0.05}$**

| Atom | X | Y | Z | Biso | Occ | Capacity from PCI [Pressure (MPa)] |
|---|---|---|---|---|---|---|
| *Ti2d* | 0.5 | 0.749(1) | 0.25 | 0.5 | 2.000 | |
| *Fe2c* | 0 | 0.297(1) | 0.25 | 0.5 | 1.824 | |
| *Ti2c* | 0 | 0.297(1) | 0.25 | 0.5 | 0.076 | |
| *Mn2c* | 0 | 0.297(1) | 0.25 | 0.5 | 0.100 | |
| *D12a* | 0 | 0 | 0 | 1.8 | 1.7(1) | |
| *D22d* | 0.5 | 0.3 | 0.25 | 1.8 | 0.3(1) | |
| *Dtotal/2* | | | | | 1.0(1) | 1.1(1) [0.21] |

**C) TiFe$_{0.80}$Mn$_{0.10}$**

| Atom | X | Y | Z | Biso | Occ | Capacity from PCI [Pressure (MPa)] |
|---|---|---|---|---|---|---|
| *Ti2d* | 0.5 | 0.745(1) | 0.25 | 0.3 | 2.000 | |
| *Fe2c* | 0 | 0.293(1) | 0.25 | 0.3 | 1.754 | |
| *Ti2c* | 0 | 0.293(1) | 0.25 | 0.3 | 0.068 | |
| *Mn2c* | 0 | 0.293(1) | 0.25 | 0.3 | 0.180 | |
| *D12a* | 0 | 0 | 0 | 1.8 | 1.7(1) | |
| *D22d* | 0.5 | 0.3 | 0.25 | 1.8 | 0.3(1) | |
| *Dtotal/2* | | | | | 0.9(1) | 1.1(1) [0.14] |

## ***α (desorption, S.G. Pm-3m)***

**A) TiFe$_{0.90}$ - ILL**

| Atom | X | Y | Z | Biso | Occ | Capacity from PCI [Pressure (MPa)] |
|---|---|---|---|---|---|---|
| *Fe1a* | 0 | 0 | 0 | 0.1 | 0.964 | |
| *Ti1a* | 0 | 0 | 0 | 0.1 | 0.036 | |
| *Ti1b* | 0.5 | 0.5 | 0.5 | 0.1 | 1.000 | |
| *D3d* | 0.5 | 0 | 0 | 1.8 | 0.07(3) | 0.01(1) [0.03] |

**A) $TiFe_{0.90}$ - ISIS**

| Atom | X | Y | Z | Biso | Occ | Capacity from PCI [Pressure (MPa)] |
|---|---|---|---|---|---|---|
| *Fe1a* | 0 | 0 | 0 | 0.3 | 0.964 | |
| *Ti1a* | 0 | 0 | 0 | 0.3 | 0.036 | |
| *Ti1b* | 0.5 | 0.5 | 0.5 | 0.4 | 1.000 | |
| *D3d* | 0.5 | 0 | 0 | 0.4 | 0.001(1) | |

**B) $TiFe_{0.85}Mn_{0.05}$ - ILL**

| Atom | X | Y | Z | Biso | Occ | Capacity from PCI [Pressure (MPa)] |
|---|---|---|---|---|---|---|
| *Fe1a* | 0 | 0 | 0 | 0.4 | 0.912 | |
| *Ti1a* | 0 | 0 | 0 | 0.4 | 0.038 | |
| *Mn1a* | 0 | 0 | 0 | 0.4 | 0.050 | |
| *Ti1b* | 0.5 | 0.5 | 0.5 | 0.4 | 1.000 | |
| *D3d* | 0.5 | 0 | 0 | 1.8 | 0.06(2) | 0.1(1) [0.02] |

**B) $TiFe_{0.85}Mn_{0.05}$ - ISIS**

| Atom | X | Y | Z | Biso | Occ | Capacity from PCI [Pressure (MPa)] |
|---|---|---|---|---|---|---|
| *Fe1a* | 0 | 0 | 0 | 0.3 | 0.912 | |
| *Ti1a* | 0 | 0 | 0 | 0.3 | 0.038 | |
| *Mn1a* | 0 | 0 | 0 | 0.3 | 0.050 | |
| *Ti1b* | 0.5 | 0.5 | 0.5 | 0.4 | 1.000 | |
| *D3d* | 0.5 | 0 | 0 | 0.4 | 0.000(1) | |

**C) $TiFe_{0.80}Mn_{0.10}$ - ILL**

| Atom | X | Y | Z | Biso | Occ | Capacity from PCI [Pressure (MPa)] |
|---|---|---|---|---|---|---|
| *Fe1a* | 0 | 0 | 0 | 0.3 | 0.877 | |
| *Ti1a* | 0 | 0 | 0 | 0.3 | 0.034 | |
| *Mn1a* | 0 | 0 | 0 | 0.3 | 0.090 | |
| *Ti1b* | 0.5 | 0.5 | 0.5 | 0.3 | 1.000 | |
| *D3d* | 0.5 | 0 | 0 | 2.0 | 0.06(4) | 0.2(1) [0.03] |

**C) $TiFe_{0.80}Mn_{0.10}$ - ISIS**

| Atom | X | Y | Z | Biso | Occ | Capacity from PCI [Pressure (MPa)] |
|---|---|---|---|---|---|---|
| *Fe1a* | 0 | 0 | 0 | 0.4 | 0.877 | |
| *Ti1a* | 0 | 0 | 0 | 0.4 | 0.034 | |
| *Mn1a* | 0 | 0 | 0 | 0.4 | 0.090 | |
| *Ti1b* | 0.5 | 0.5 | 0.5 | 0.3 | 1.000 | |
| *D3d* | 0.5 | 0 | 0 | 0.3 | 0.003(1) | |

## *β (absorption, S.G. $P222_1$)*

**A) $TiFe_{0.90}$**

| Atom | X | Y | Z | Biso | Occ | Capacity from PCI [Pressure (MPa)] |
|---|---|---|---|---|---|---|
| *Ti2d* | 0.5 | 0.747(3) | 0.25 | 0.1 | 2.000 | |
| *Fe2c* | 0 | 0.291(1) | 0.25 | 0.1 | 1.928 | |
| *Ti2c* | 0 | 0.291(1) | 0.25 | 0.1 | 0.072 | |
| *D12a* | 0 | 0 | 0 | 1.8 | 2.1(1) | |
| *D22d* | 0.5 | 0.3 | 0.25 | 1.8 | 1.0(1) | |
| *Dtotal/2* | | | | | 1.5(1) | 1.2(1) [1.78] |

**B) $TiFe_{0.85}Mn_{0.05}$**

| Atom | X | Y | Z | Biso | Occ | Capacity from PCI [Pressure (MPa)] |
|---|---|---|---|---|---|---|
| *Ti2d* | 0.5 | 0.749(1) | 0.25 | 0.5 | 2.000 | |
| *Fe2c* | 0 | 0.297(1) | 0.25 | 0.5 | 1.824 | |
| *Ti2c* | 0 | 0.297(1) | 0.25 | 0.5 | 0.076 | |
| *Mn2c* | 0 | 0.297(1) | 0.25 | 0.5 | 0.100 | |
| *D12a* | 0 | 0 | 0 | 1.8 | 1.9(1) | |
| *D22d* | 0.5 | 0.3 | 0.25 | 1.8 | 0.7(1) | |
| *Dtotal/2* | | | | | 1.3(1) | 1.3(1) [0.71] |

## *γ ($2^{nd}$ absorption, S.G. Cmmm)*

**B) $TiFe_{0.85}Mn_{0.05}$ (D1B)**

| Atom | X | Y | Z | Biso | Occ | Capacity from PCI [Pressure (MPa)] |
|---|---|---|---|---|---|---|
| *Ti4h* | 0.213(1) | 0 | 0.5 | 0.4 | 4.000 | |
| *Fe4i* | 0 | 0.286(1) | 0 | 0.4 | 3.648 | |
| *Ti4i* | 0 | 0.286(1) | 0 | 0.4 | 0.152 | |
| *Mn4i* | 0 | 0.286(1) | 0 | 0.4 | 0.200 | |
| *D14e* | 0.25 | 0.25 | 0 | 2.0 | 4.6(1) | |
| *D12c* | 0.5 | 0 | 0.5 | 2.0 | 1.6(1) | |
| *D12a* | 0 | 0 | 0 | 2.0 | 1.8(1) | |
| *Dtotal/4* | | | | | 2.0(1) | 1.9(1) [4.90] |

**B) $TiFe_{0.85}Mn_{0.05}$ (D2B)**

| Atom | X | Y | Z | Biso | Occ | Capacity from PCI [Pressure (MPa)] |
|---|---|---|---|---|---|---|
| *Ti4h* | 0.220(1) | 0 | 0.5 | 0.4 | 4.000 | |
| *Fe4i* | 0 | 0.285(1) | 0 | 0.4 | 3.648 | |
| *Ti4i* | 0 | 0.285(1) | 0 | 0.4 | 0.152 | |
| *Mn4i* | 0 | 0.285(1) | 0 | 0.4 | 0.200 | |
| *D14e* | 0.25 | 0.25 | 0 | 2.0 | 4.2(1) | |
| *D12c* | 0.5 | 0 | 0.5 | 2.0 | 1.8(1) | |
| *D12a* | 0 | 0 | 0 | 2.0 | 1.7(1) | |
| *Dtotal/4* | | | | | 1.8(1) | |